\documentclass[prl,twocolumn,superscriptaddress,floatfix]{revtex4}

       \usepackage{epsf}
       \usepackage{graphicx}
       \usepackage{dcolumn}
       \usepackage{bm}

       \newcommand{\be}{\begin{equation}}
       \newcommand{\ee}{\end{equation}}
       \newcommand{\ba}{\begin{eqnarray}}
       \newcommand{\ea}{\end{eqnarray}}

\begin{document}

       \title{Landau-Zener Transitions in an Adiabatic Quantum Computer}

       \author{J. Johansson}\email{jjohansson@dwavesys.com} \author{M.H.S. Amin} \author{A.J. Berkley} \author{P. Bunyk}
       \author{V. Choi} \author{R. Harris} \author{M.W. Johnson} \author{T.M. Lanting}
       \affiliation{D-Wave Systems Inc., 100-4401 Still Creek Drive,
       Burnaby, B.C., V5C 6G9, Canada}
       \author{Seth Lloyd}
       \affiliation{W.M.Keck Center for Extreme Quantum Information Processing
       (xQIT), MIT3-160, Cambridge, MA 02139 USA}
        \author{G. Rose}
        \affiliation{D-Wave Systems Inc., 100-4401 Still Creek Drive,
       Burnaby, B.C., V5C 6G9, Canada}

\begin{abstract}
We report an experimental measurement of Landau-Zener transitions on
an individual flux qubit within a multi-qubit superconducting chip
designed for adiabatic quantum computation. The method used isolates
a single qubit, tunes its tunneling amplitude $\Delta$ into the
limit where $\Delta$ is much less than both the temperature $T$ and
the decoherence-induced energy level broadening, and forces it to
undergo a Landau-Zener transition. We find that the behavior of the
qubit agrees to a high degree of accuracy with theoretical
predictions for Landau-Zener transition probabilities for a
double-well quantum system coupled to 1/f magnetic flux noise.
\end{abstract}

\maketitle

Adiabatic quantum computation (AQC) is a quantum mechanical method
for solving hard computational problems \cite{Farhi}. In AQC one
encodes a computational problem in a suitable physical system. If
one can somehow put the system in its ground, or lowest energy
state, the structure of that ground state then reveals the answer to
the problem. To find that ground state using AQC, one starts by
engineering a simple Hamiltonian or energy functional for the
system, and by placing the system in the ground state of this simple
Hamiltonian.  Then one gradually deforms the Hamiltonian of the
system from the simple form into a complex Hamiltonian whose ground
state encodes the answer to the problem.  If this deformation is
sufficiently gradual, then the transformation of the state of the
system is adiabatic, and the system remains in its ground state
throughout the deformation. AQC is known to be a universal model of
quantum computation \cite{Aharonov07}.

A key question of AQC is whether adiabaticity can be maintained
throughout the computation. For an isolated system, Landau-Zener
(LZ) transitions \cite{Landau,Zener} may take the system out of its
ground state, unless the time over which one transverses the point
of the minimum energy gap during the course of a computation is
longer than the instantaneous coupling between ground and excited
states divided by the minimum gap squared. Just which hard problems
can be encoded in such a way that adiabaticity can be maintained
over reasonable times is an open question. Moreover, for some
optimization problems, an excited final state with sufficiently low
energy could provide an acceptable solution and therefore the
adiabaticity condition may be relaxed.

Equally important question is whether interactions between the
computer and its environment can spoil the computation. Unlike in
the gate model of quantum computation \cite{Ike}, the effects of an
environment on AQC are less understood. While it is now clear that
AQC has fundamental advantages over the gate model in regards to
robustness against decoherence \cite{Childs,ALT08,AA07,ATA08}, there
does not yet exist an equivalent of the threshold theorem \cite{Ike}
that describes under what conditions AQC coupled to an environment
will succeed. Nonetheless, a clear understanding of how LZ
transitions are affected by an environment represents an important
step forward.

For large-scale hard problems, it is inevitable that any
implementation of AQC will encounter minimum gaps that are smaller
than temperature $T$ and decoherence rate. In order to understand
the effects of environment in this limit, we attempt to first
understand in detail how environment affects LZ transitions in
individual qubits within an adiabatic quantum computer. We isolate a
single qubit in a 28 qubit superconducting chip from its surrounding
qubits, and measure its behavior as it undergoes a LZ transition. We
purposely operate in a regime in which the decoherence time scale
$\tau_\varphi$ is much shorter than the adiabatic passage, and in
which $T \gg \Delta$, where $\Delta$ is the tunneling amplitude.
While evidence of LZ transitions in superconducting qubits has been
reported before \cite{Izmalkov,Oliver,Sillanpaa}, to our knowledge
this is the first {\em direct} measurement of LZ transitions in a
superconducting qubit in the high $T$ and strong decoherence regime.

In the original LZ problem, the system Hamiltonian is
       \begin{equation}
       H_S = -(\Delta \sigma_x + \epsilon \sigma_z)/2, \label{HS}
       \end{equation}
with $\epsilon = \nu t$, where $\sigma_{x,z}$ are Pauli matrices and
$\nu$ is the sweep rate for the energy bias. We take $|0\rangle$ and
$|1\rangle$ to be eigenfunctions of $\sigma_z$, denoting the ``left"
and ``right" states in a double-well potential which can represent
the two flux states in a superconducting flux qubit. If at
$t=-\infty$ the system starts in state $|0\rangle$, then the
probability of finding the system in same state at time $t=+\infty$
is {\em exactly} given by \cite{Landau,Zener}
 \be
 P_{\rm LZ} = e^{-\pi \Delta^2/2\nu}. \label{LZEq}
 \ee
If Hamiltonian (\ref{HS}) describes the dynamics of the two lowest
energy states in a multi-qubit adiabatic quantum computer close to
the energy anticrossing, then (\ref{LZEq}) is the probability of
failing to reach the final ground state in the decoherence-free
system.

Now suppose that the qubit is coupled to an environment. The total
Hamiltonian $H=H_S+H_B+H_{\rm int}$, comprises the system (\ref{HS})
and environment $H_B$ parts, and an interaction Hamiltonian
\begin{equation}
  H_{\rm int} = -Q \sigma_z/2, \label{Hint}
\end{equation}
that provides coupling between the qubit and an operator $Q$ that
acts on the environment. Here, we only consider longitudinal
coupling to the environment which represents flux noise affecting
the flux bias in a flux qubit. We don't specify $H_B$ explicitly,
because if environmental fluctuations obey Gaussian statistics, then
all averages can be expressed in terms of the spectral density
$S(\omega) = \int_{-\infty}^\infty dt \
e^{i\omega t}\langle Q(t)Q(0)\rangle$.
Here, $\langle ... \rangle$ denotes averaging over environmental
degrees of freedom. Hamiltonian (\ref{Hint}) is what one expects for
the effective interaction Hamiltonian for a large-scale AQC at the
anticrossing, regardless of the type of coupling of individual
qubits to the environment \cite{ATA08}.

An immediate consequence of coupling to the environment is that the
relative phase between the two terms in the wave function that
correspond to the two energy levels becomes uncertain after some
time, an effect known as dephasing or decoherence. Due to
energy-time uncertainty, the dephasing time $\tau_\varphi$ is
related to the uncertainty in the energy eigenvalues or so called
broadening $W$ of the energy levels: $1/\tau_\varphi=W
|\epsilon|/\sqrt{\epsilon^2+\Delta^2}$. If $W,T \ll \Delta$, then
the system will have a well-defined ground state separated from the
excited state by a well-defined gap and thermal transitions will be
suppressed. One would then expect that (\ref{LZEq}) holds even in
the presence of noise, although $\Delta$ may be renormalized by high
frequency modes of the environment \cite{Leggett,Weiss}. The
important question is now what happens when $W,T \gg \Delta$ so that
the broadened ground and first excited states merge into each other
and thermal transitions completely mix them up. Here, we answer this
question both theoretically and experimentally.

In the regime $\Delta \ll W$, the dynamics of the system become
incoherent. Using second order perturbation in $\Delta$ and assuming
that the environmental is dominated by low frequency Gaussian noise,
the incoherent tunneling rate from $|0\rangle$ to $|1\rangle$ is
given by \cite{MRTtheory}
 
\begin{eqnarray}
  \Gamma_{01} (\epsilon) = \sqrt{\pi \over 8}{\Delta^2 \over W} \exp
  \left\{-{(\epsilon - \epsilon_p)^2 \over 2W^2}\right\}, \ \
  \label{Gamma} \\
  W^2 = \int {d\omega \over 2\pi} S(\omega) \; \; , \; \;  \epsilon_p = {\cal P} \int
  {d\omega \over 2\pi}{S(\omega) \over \omega}, \label{Wep}
\end{eqnarray}

with backward transition given by
$\Gamma_{10}(\epsilon)=\Gamma_{01}(-\epsilon)$. The transition rates
therefore exhibit a Gaussian peak with a center shifted away from
the resonance point $\epsilon=0$. The width of the transition
region, $W$, which is a measure of the environmentally induced
broadening of the energy levels, is thus given by the r.m.s. value
of the noise. In thermal equilibrium, the width $W$ and the position
$\epsilon_p$ of such macroscopic resonant tunneling (MRT) peak are
related by \cite{MRTtheory}
       \begin{equation} W^2 = 2T  \epsilon_p, \label{equilb} \end{equation}
These predictions have been experimentally confirmed using
superconducting flux qubits \cite{Harris07}.

Let us now return to the LZ problem. Suppose at $t=t_i$ the system
starts from $|0\rangle$ with probability $P_0(t_i)=1$. In the
incoherent tunneling regime ($\Delta \ll W$), the off-diagonal
elements of the density matrix vanish quickly (within time scale
$\tau_\varphi\sim 1/W$). Thus to find $P_0(t)$, one needs to solve
the equation
 \be
 \dot P_0 = -\Gamma_{01} P_0 + \Gamma_{10} (1-P_0). \label{masterEq}
 \ee
In the low temperature regime $T\ll W$ (but can be $\gg \Delta$),
Eq.~(\ref{equilb}) requires that $W\ll 2\epsilon_p$, thus separating
the peaks of $\Gamma_{01}$ and $\Gamma_{10}$ such that
$\Gamma_{01}(-\epsilon_p) \ll \Gamma_{01}(\epsilon_p)$. One can
therefore neglect the second term in (\ref{masterEq}); because at
points where $\Gamma_{10}$ is peaked, $1-P_0 \approx 0$, and at all
other points, $\Gamma_{10}\approx 0$. The probability $P_0$ will
then be approximately given by
 \ba
 P_0(t_f) = e^{-\int_{t_i}^{t_f} \Gamma_{01}(\tau)d\tau}=e^{-(\pi \Delta^2/2\nu)
 \kappa}, \label{P0f} \\
 \kappa = {1\over \sqrt{2\pi}W} \int_{\epsilon_i}^{\epsilon_f}
 \exp \left\{-{(\epsilon - \epsilon_p)^2 \over 2W^2}\right\}
 d\epsilon.
 \ea
If $\epsilon_i \to -\infty$, this equation becomes
 \be
 \kappa = {1\over 2} \left[1+
 \text{erf} \left( {\epsilon_f - \epsilon_p \over \sqrt{2}W}\right)
 \right]. \label{kappa}
 \ee
If also $\epsilon_f \to \infty$, then $\kappa = 1$, yielding
(\ref{LZEq}), which is exactly the LZ transition probability in a
completely coherent system, in agreement with previous studies
\cite{Ao}. If the condition $W \gg T$ does not hold, then one must
keep all of the terms in (\ref{masterEq}) and calculate $P_0$
numerically.

\begin{figure}[t]
\includegraphics[width=8.5cm]{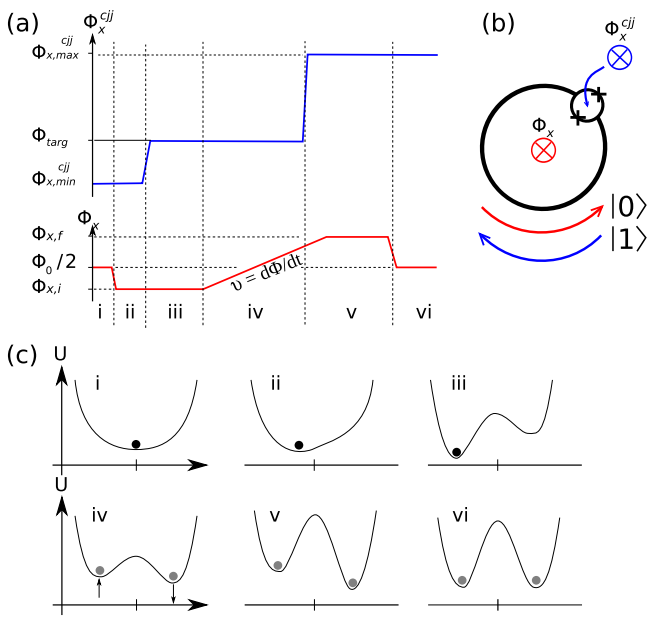}\\
\caption{Schematics of (a) the pulse sequence for the LZ
measurements, (b) a CJJ RF-SQUID qubit, and (c) the qubit potential
during the measurement.} \label{fig1}
\end{figure}

We have experimentally tested the above predictions by examining LZ
transitions using a single qubit in a 28 qubit chip designed for
adiabatic quantum computation. The sample was cooled down in a
magnetically shielded dilution refrigerator with heavily filtered
lines to a base temperature of about 10 mK. The qubits on the chip
were compound Josephson junction (CJJ) RF-SQUID qubits as
schematically shown in Fig.~\ref{fig1}b and described in
Ref.~\onlinecite{Harris07}. Each qubit consists of a main loop and
CJJ loop subjected to external flux biases $\Phi_x$ and
$\Phi_x^{\text{cjj}}$, respectively. The CJJ loop is interrupted by
two nominally identical Josephson junctions connected in parallel.
This device can be operated as a qubit for $\Phi_{x}^{\text{cjj}}\in
\lbrack 0.5,1\rbrack\Phi _{0}$ and $\Phi _{x}\approx 0$, where
$\Phi_0$ is the flux quantum. The two oppositely circulating
persistent current states correspond to the states $\left|0\right>$
and $\left|1\right>$. The bias energy is $\epsilon =2\left\vert
I_{p}\right\vert\Phi _{x}$, where the $I_{p}$ is the magnitude of
the persistent current. The parameter $\Delta$ is the amplitude of
the flux tunneling between the two states. Both $\left|I_p\right|$
and $\Delta$ are controlled by $\Phi_{x}^{\text{cjj}}$. Maximum
$\Delta \sim\omega_p \sim 20$ GHz, where $\omega_p$ is the plasma
frequency of the RF-SQUID, is obtained at $\Phi
_{x}^{\text{cjj}}=\Phi_{0}/2$. For $\Phi
_{x}^{\text{cjj}}\approx\Phi_0$ one expects $\Delta \rightarrow 0$
and the system becomes localized in $\left|0\right>$ or
$\left|1\right>$. One can then read out the qubit by measuring the
flux via an inductively coupled DC-SQUID (not shown).

\begin{figure}[t]
\includegraphics[width=8.5cm]{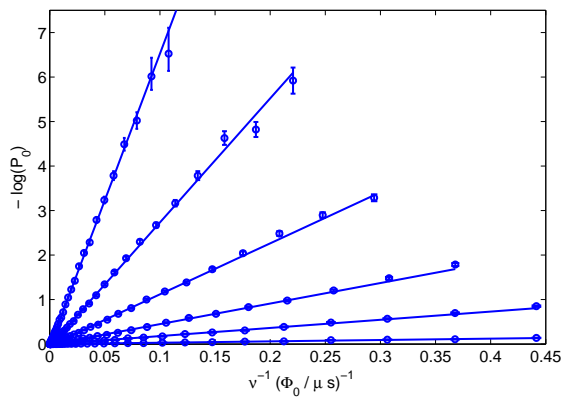}
\caption{LZ probability as a function of inverse sweep rate for
different values of $\Delta$ ($\Phi_x^{cjj}$ from  $-0.925 \Phi_0$
to  $-0.91 \Phi_0$). The lines are linear fits to the
data.}\label{fig2}
\end{figure}

We isolated one of the qubits by tuning the coupling between that
qubit and its neighboring qubits to zero, which allowed us to
perform single qubit LZ measurement using the pulse sequence shown
in Fig.~\ref{fig1}a. The qubit is first initialized in one of the
states $|0\rangle$ or $|1\rangle$ at a bias $\Phi_x = \Phi_{x,i}$
where the transition out of the state is very unlikely, and then the
bias is linearly swept from $\Phi_{x,i}$ to a final value
$\Phi_{x,f}$, at which point the qubit is measured. A cartoon of the
qubit potential during the pulse sequence is shown in
Fig.~\ref{fig1}c. The probability $P_0(t_f)$ of finding the qubit in
the same state $|0\rangle$ as it started from was measured by
repeating the above process 2048 times for each value of the sweep
rate $\nu$. Figure~\ref{fig2} shows the probability $P_0(t_f)$ in
logarithmic scale as a function of $1/\nu$. The result shows
exponential dependence upon $1/\nu$, in agreement with
Eq.~(\ref{P0f}).

\begin{figure}[t]
\includegraphics[width=9cm]{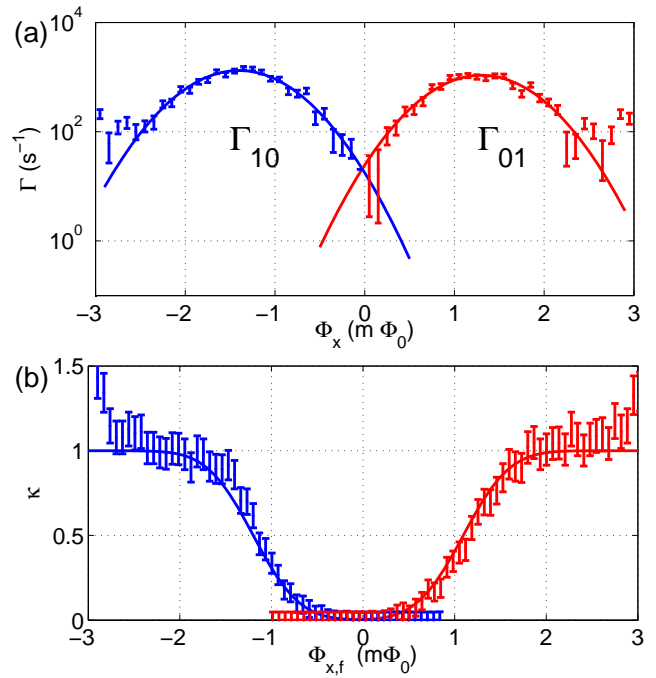}
\caption{(a) First MRT peaks in $\Gamma_{01}$ and $\Gamma_{10}$ and
their best fit with Eq.~(\ref{Gamma}). (b) The experimental value of
$\kappa$ as defined in (\ref{P0f}) as a function of the final bias
$\epsilon_f$. The solid line shows the theoretical curve
(\ref{kappa}) using the parameters obtained from the best fit in
(a).}\label{fig3}
\end{figure}

Next we experimentally verify equation (\ref{kappa}). We first
determine $\Delta$, $W$ and $\epsilon_p$ by measuring $\Gamma_{01}$
and $\Gamma_{10}$, as described in Ref.~\onlinecite{Harris07}.
Figure~\ref{fig3}a shows example plots of $\Gamma_{01}$ and
$\Gamma_{10}$ as a function of bias $\Phi_x$ for the above qubit at
$\Phi_x^{\text{cjj}} = -0.749\ \Phi_0$. The line-shape of the
resonant peak fits very well with the Gaussian function
(\ref{Gamma}), providing $\Delta, W$, and $\epsilon_p$ as fitting
parameters. For the data shown in Fig.~\ref{fig3}a, we found $\Delta
= 0.082 \pm 0.002$ mK, $W = 123 \pm 2$ mK, and $\epsilon_p = 354 \pm
3$ mK. Equation (\ref{equilb}) then gives the effective temperature
of the sample to be $T=21 \pm 1$ mK. Notice that the condition $T
\ll W \ll 2\epsilon_p$ is approximately satisfied and therefore
(\ref{P0f}) should be sufficient to describe the LZ probability.

The LZ probability as a function of flux bias was then measured for
the same CJJ setting as in Fig.~\ref{fig3}a. Figure~\ref{fig3}b
shows $\kappa=-\ln P_0/(\pi \Delta / 2\nu)$ as a function of
$\Phi_{x,f}$ using the extracted $\Delta$. The data starts from zero
where $\Phi_{x,f} \approx \Phi_{x,i}$ and shows a plateau at $\kappa
\approx 1$, in agreement with the theory. We have also plotted, on
the same graph, the theoretical prediction of Eq.~(\ref{kappa}),
using $\epsilon_p$ and $W$ extracted from the MRT in
Fig.~\ref{fig2}a, and found very good agreement with the experiment
with no extra fitting parameters. At larger biases, $| \Phi_{x,f} |
> 2.5\ \mathrm{m}\Phi_0$, the experimental $\kappa$ deviates from
the theoretical curve due to tunneling to the first excited state in
the target well.


\begin{figure}[t]
\includegraphics[width=8.5cm]{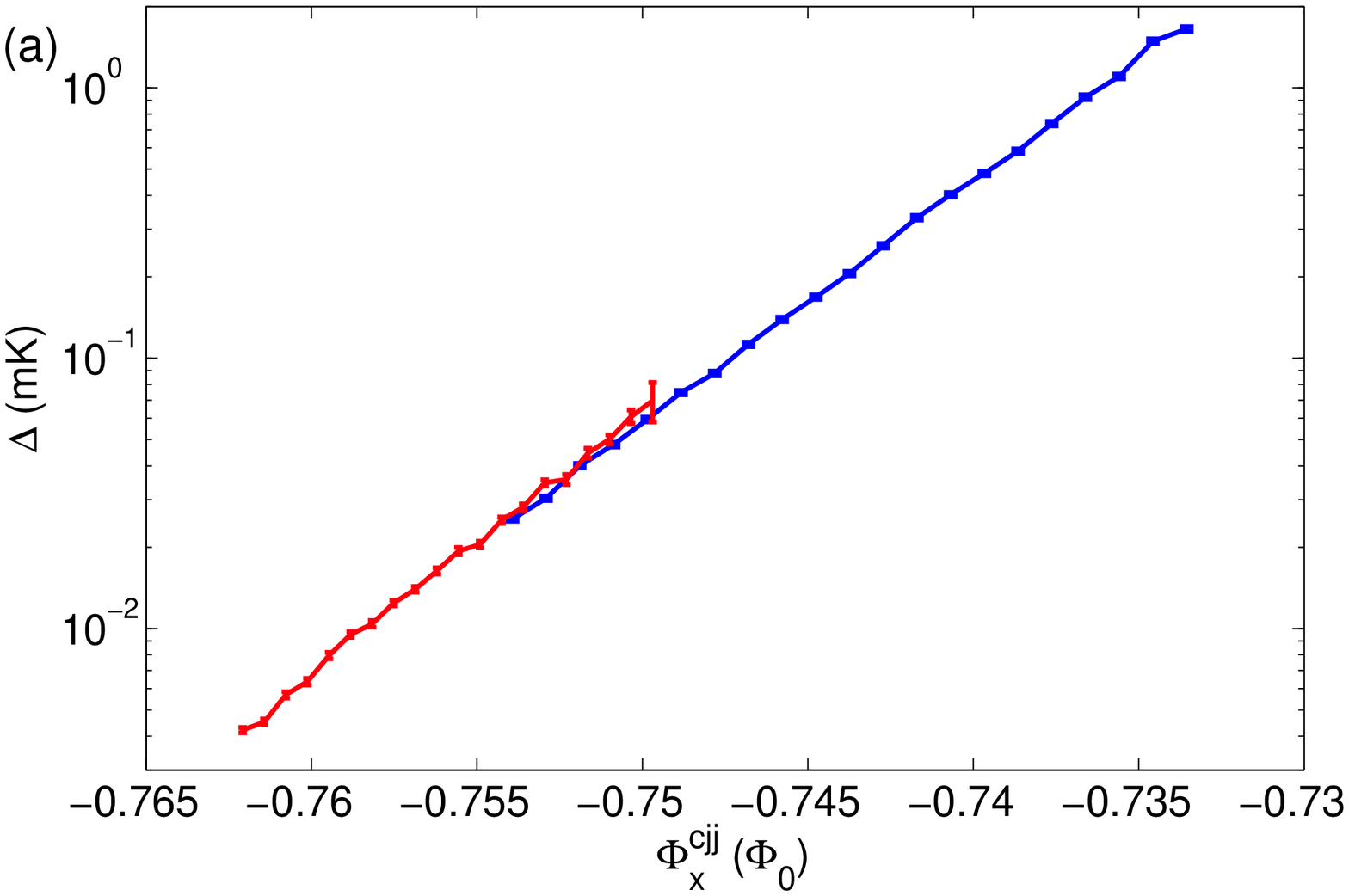}
\includegraphics[width=8.5cm]{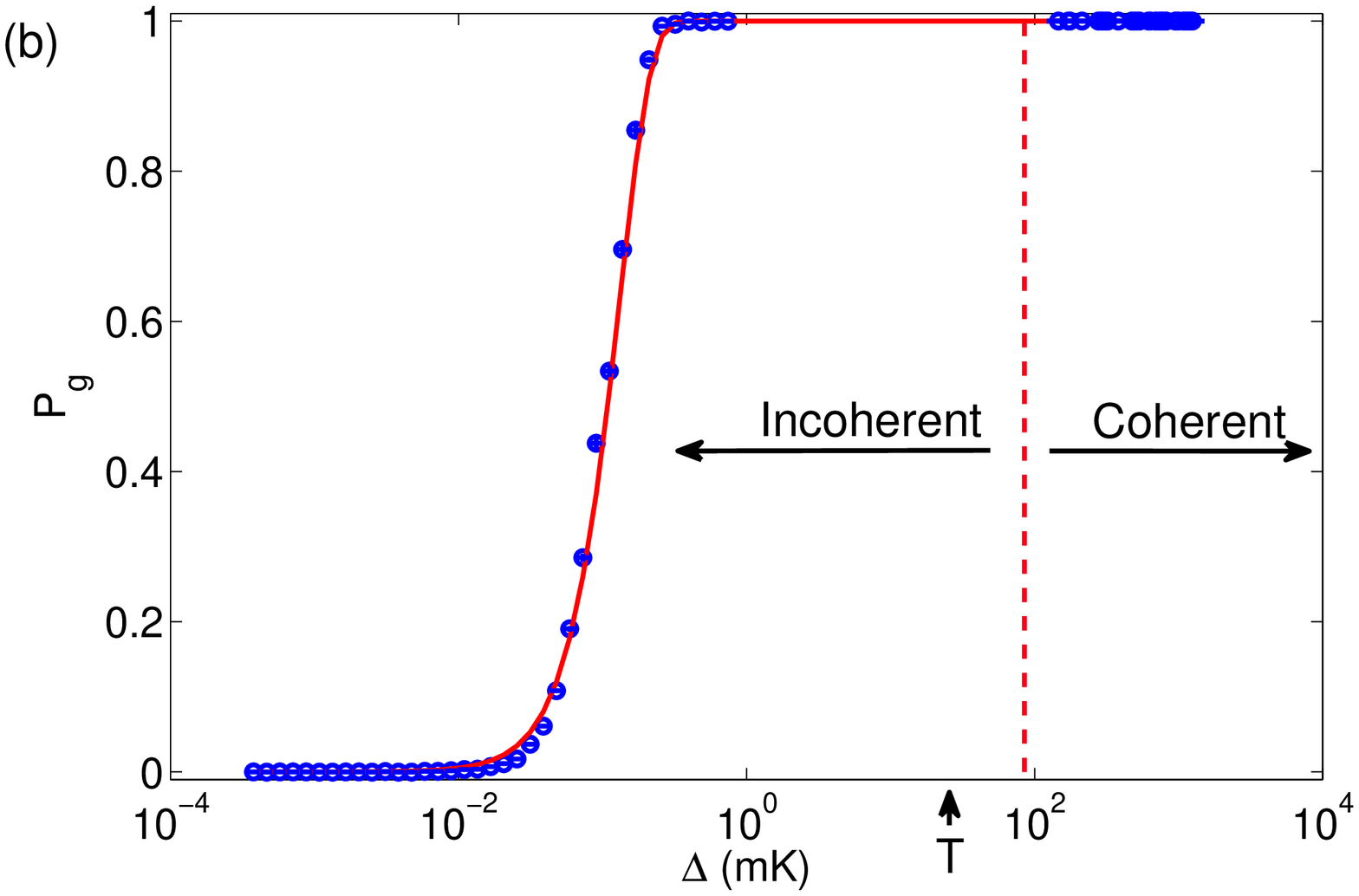}
\caption{(a) Plot of $\Delta$ vs. flux bias $\Phi_x^{cjj}$ applied
to the compound junction. The lower curve is $\Delta$ measured with
MRT and the upper curve is $\Delta$ measured with LZ. (b)
Experimental (dots) and theoretical (line) ground state probability
$P_g=1-P_{\rm LZ}$, for fixed $\nu = 0.05 \Phi_0/\mu s$ as a
function of $\Delta$. The dashed line indicates the crossover
between the incoherent and coherent regime, defined by $W \approx
\Delta$.}\label{fig4}
\end{figure}

The measurements of the transition rates and the LZ probability
allow us to extract $\Delta$ as a function  $\Phi_x^{\text{cjj}}$
for a large range of $\Delta$ \footnote{The range is limited from
below and above by longest practical measurement time and bandwidth
of the measurement lines respectively.}. Exponential dependence on
$\Phi_x^{\text{cjj}}$ is evident in Fig.~\ref{fig4}a.
Figure~\ref{fig4}b plots the LZ probability, for a fixed value of
$\nu$, as a function of $\Delta$ for a quite wide range of $\Delta$
(from 27 $\mu$K to 1.25 K) together with the theoretical prediction
\footnote{$\Delta$ in the coherent regime was obtained by measuring
the average flux and using $\langle \Phi \rangle \propto \langle
\sigma_z\rangle = \epsilon/\sqrt{\epsilon^2+\Delta^2}$, which is
valid in the large gap regime.}. In the figure we have identified a
line $\Delta=W$ which separates coherent tunneling from incoherent
tunneling regime. Excellent agreement with theory is observed.

We have reported on an experimental probe of the practically
interesting regime for adiabatic quantum computation, where the
energy gap $\Delta$ is much smaller than both the decoherence
induced energy level broadening $W$ and temperature $T$. The method
used isolates a single qubit in a larger-scale adiabatic quantum
computer, tunes its tunneling amplitude $\Delta$ into the $\Delta
\ll W, T$ limit, and forces it to undergo a LZ transition. We find
that the transition probability for the qubit quantitatively agrees
with the theoretical predictions. In particular, we demonstrate that
in this large decoherence limit, the quantum mechanical behavior of
this qubit is the same (except for possible renormalization of
$\Delta$) as that of a noise-free qubit, as long as the energy bias
sweep covers the entire region of broadening $W$. The close
agreement between theory and experiment for a single qubit
undergoing a LZ transition in the presence of noise supports the
accuracy of our dynamical models, including both the noise model and
the model of a single superconducting qubit that has been isolated
from its surrounding qubits in an adiabatic quantum computer. Future
experiments will test the behavior of multiple coupled qubits
undergoing a LZ transition in the presence of noise.

The authors are grateful to D.V.~Averin for fruitful discussions.
Samples were fabricated by the Microelectronics Laboratory of the
Jet Propulsion Laboratory, operated by the California Institute of
Technology under a contract with NASA.

       \end{document}